\documentclass[aps,prl,floats,twocolumn,showpacs,superscriptaddress]{revtex4}

\usepackage{epsfig}
\usepackage{latexsym,amsmath}
\usepackage{amsbsy}

\begin{document}
\title{Correlations in weighted networks}

\author{M. \'Angeles Serrano}
\affiliation{School of Informatics, Indiana University, Eigenmann
Hall, 1900 East Tenth Street, Bloomington, IN 47406, USA}
\affiliation{Complex Network Lagrange Laboratory (CNLL), Institute for Scientific Interchange (ISI), Torino, Italy}

\author{Mari{\'a}n Bogu{\~n}{\'a}}
\affiliation{Departament de F{\'\i}sica Fonamental, Universitat de
  Barcelona,\\ Mart\'{\i} i Franqu\`es 1, 08028 Barcelona, Spain}

\author{Romualdo Pastor-Satorras}

\affiliation{Departament de F{\'\i}sica i Enginyeria Nuclear, Universitat
  Polit{\`e}cnica de Catalunya, Campus Nord, M{\'o}dul  B4, 08034 Barcelona, Spain}

\date{\today}

\begin{abstract}
We develop a statistical theory to characterize correlations in weighted
networks. We define the appropriate metrics quantifying correlations and show
that strictly uncorrelated weighted networks do not exist due to the presence
of structural constraints. We also introduce an algorithm for generating
maximally random weighted networks with arbitrary $P(k,s)$ to be used as null
models. The application of our measures to real networks reveals the
importance of weights in a correct understanding and modeling of these
heterogeneous systems.
\end{abstract}


\maketitle

In the current era of fast technological progress, heterogeneous
transport systems appear at the core of the last revolutionary
advances. The information technology revolution represents maybe one
of the most outstanding examples, with the
Internet~\cite{RomusVespasbook} factually reshaping the ways of social
and economic interactions. The success of this revolution is, at the
same time, intimately linked to the development of other infrastructures also
involving transference. This is the case of the globalized
transportation systems and, in particular, of the worldwide airport
network~\cite{Barrat:2004b,mossa2}, which serves as a ground for the
transport of people, goods, and even diseases~\cite{Colizza:2006}
throughout the world in a very short time scale. Due to their profound
and far-reaching impact, it is crucial to develop theoretical tools
to increse our understanding of the large scale properties of these
systems, which can help to take actions in their engineering
against possible malfunction or jamming.

Both the Internet and the worldwide air transportation system, and in
general most heterogeneous transport systems, can be represented as
weighted complex networks (WCNs) \cite{Dorogovtsev:2003}, in which
vertices stand for the elementary units composing the system and edges
represent the interactions or relations between pairs of units. The
latter are further characterized by a weight measuring
the capacity or the amount of traffic in a particular
connection~\cite{Clark:1990}.  Although the theory of unweighted
complex networks, where edges are exclusively modulated as present or
absent, is today well
established~\cite{Albert:2002,Dorogovtsev:2003,Boguna:2003b},
there is not yet available an equivalent formalism for the weighted
case and the present knowledge comes from particular models of growing
WCNs \cite{Yook:2001,barrat04:_weigh}.  This
makes difficult to define suitable observables to characterize these systems properly.  For instance, several definitions of
the basic correlation functions~\cite{Boguna:2003b} have been
suggested~\cite{Barrat:2004b,Onnela:2005,Ahnert:2006}, but it is not
clear which of those provide the correct measures. And what is worse,
no proper null model for the presence of correlations has been
proposed in order to compare with empirical data. Null models are particularly relevant in this context because heterogeneous networks usually display unavoidable structural correlations which can lead to a mistaken understanding of the principles that shape the system and its functionality~\cite{Boguna:2004,Colizza:2006b}.

In this paper, we fill this gap by introducing a rigorous framework for
the characterization of correlations in WCNs that allows to define proper measures. We shall see that, at the weighted level, strictly uncorrelated networks do not exists due to structural constraints. Yet, our formalism enables to define an algorithm that generates maximally random WCNs with arbitrary local properties to be used
as a null model with respect non-structural correlations. This algorithm
corresponds to a weighted version of the random graph ensemble
proposed by Chung and Lu~\cite{Chung:2002}. We also define correlation measures that filter out the structural constraints. As an example, we apply
our formalism to the US airport system~\cite{USANData} (USAN), the scientific collaboration network~\cite{Newman:2001a} (SCN), and the world trade web~\cite{Serrano:2003} (WTW). The information obtained reveals that weights, rather than the bare topology, rule the architecture of some of them.   

Unweighted networks can be fully characterized by means of a binary
variable $a_{ij}$, taking the values $a_{ij}=1$ when the edge between
vertices $i$ and $j$ is present and $0$ otherwise. Relevant
statistical topological properties can then be derived from this
adjacency matrix, more specifically, the degree distribution $P(k)$,
defined as the probability that a vertex is connected to $k$ other
vertices, or degree correlations measured by the average degree of the
nearest neighbors as a function of the vertex degree,
$\bar{k}_{nn}(k)$~\cite{Pastor-Satorras:2001}, and the
degree-dependent clustering coefficient
$\bar{c}(k)$~\cite{Watts:1998,Vazquez:2002}. In the case of WCNs,
edges have assigned a real or natural number $w_{ij}$, representing
the weight or intensity of the connection between $i$ and $j$. Thus,
apart from the vertex degree $k_i$, the presence of weights allows to
define other significant properties, such as the vertex strength
$s_i$~\cite{Barrat:2004b,Yook:2001}, given by $s_i = \sum_j w_{ij}$,
and statistical distributions such as the strength distribution $P(s)$, the average
strength of vertices of degree $k$, $\bar{s}(k)$, or, in a more
general way, the joint probability $P(k,s)$ that a vertex
has degree $k$ and strength $s$, simultaneously. However, the strength alone is not
enough to capture the weighted structure of vertices since the ratio
$s/k$ gives only the average weight per connection but says nothing
about fluctuations around this average. Therefore, we need to
introduce some measure of the \textit{fluctuations} of weights of a
given vertex. To this end, we use the \textit{disparity} $Y$, defined
as $Y_{i} =\sum_{j} (w_{ij}/s_{i})^2$~\cite{Barthelemy:2005}.
Now, our main hypothesis is that all vertices with the same degree,
strength, and disparity, that is, characterized by the same vector
variable $\boldsymbol{\alpha}=(k,s,Y)$, are statistically equivalent, so that we can define 
$P(\boldsymbol{\alpha}) \equiv P(k,s,Y)$ as the probability that a given
vertex has degree $k$, strength $s$, and disparity $Y$.
Without lack of generality, we will also assume that the strength is a
discrete variable so that the equivalence classes form a numerable
set.

To quantify two-point correlations for weighted networks, we start by defining
two matrices~\cite{Boguna:2003b}. Let $E_{\boldsymbol{\alpha},
\boldsymbol{\alpha}'}$ be the matrix accounting for the number of connections
between the class of vertices $\boldsymbol{\alpha}$ and the class of vertices
$\boldsymbol{\alpha}'$ (two times this number if the two classes are the
same). Analogously, let $W_{\boldsymbol{\alpha}, \boldsymbol{\alpha}'}$ be the
matrix that accounts for the weight between the same pair of classes. Let $N$,
$E$, and $W$ be the number of vertices, edges, and total weight of the
network, respectively. Then, the \textit{fundamental functions} characterizing
the two-point correlation structure in WCNs are
\begin{equation}
  P(\boldsymbol{\alpha}, \boldsymbol{\alpha}') \equiv
  \frac{E_{\boldsymbol{\alpha},\boldsymbol{\alpha}'}}{\langle k \rangle N}
  \quad \mathrm{and} \quad
  Q(\boldsymbol{\alpha}, \boldsymbol{\alpha}') \equiv
  \frac{W_{\boldsymbol{\alpha}, \boldsymbol{\alpha}'}}{\langle s \rangle N}.
\label{eq:1}
\end{equation}
Both functions have a clear interpretation~\cite{Boguna:2003b}. Indeed,
$(2-\delta_{\boldsymbol{\alpha}, \boldsymbol{\alpha}'})P(\boldsymbol{\alpha},
\boldsymbol{\alpha}')$ is the probability that a randomly chosen edge of the
network connects two vertices of the classes $\boldsymbol{\alpha}$ and
$\boldsymbol{\alpha}'$. Analogously, $(2-\delta_{\boldsymbol{\alpha},
\boldsymbol{\alpha}'})Q(\boldsymbol{\alpha}, \boldsymbol{\alpha}')$ gives the
probability that, when choosing an edge of the network with a probability
proportional to its weight, this edge connects two vertices of the classes
$\boldsymbol{\alpha}$ and $\boldsymbol{\alpha}'$. These fundamental functions
satisfy the summation rules $\sum_{\boldsymbol{\alpha}'}
P(\boldsymbol{\alpha},\boldsymbol{\alpha}')=kP(\boldsymbol{\alpha})/\langle k
\rangle$ and $\sum_{\boldsymbol{\alpha}'}
Q(\boldsymbol{\alpha},\boldsymbol{\alpha}')=sP(\boldsymbol{\alpha})/\langle s
\rangle$. This allow to define the relevant conditional probabilities
\begin{equation}
  P(\boldsymbol{\alpha}' | \boldsymbol{\alpha}) =  \frac{\langle k \rangle
      P(\boldsymbol{\alpha}, \boldsymbol{\alpha}')}{k P(\boldsymbol{\alpha})}
      \mbox{ \hspace{0.05cm}, \hspace{0.05cm}} 
        Q(\boldsymbol{\alpha}' | \boldsymbol{\alpha})  = \frac{\langle s \rangle
    Q(\boldsymbol{\alpha}, \boldsymbol{\alpha}')}{s P(\boldsymbol{\alpha})}.
\label{eq:2}
\end{equation}
As usual, $P(\boldsymbol{\alpha}' | \boldsymbol{\alpha})$ measures the
probability that a randomly chosen edge from a vertex in the class
$\boldsymbol{\alpha}$ points to a vertex in the class $\boldsymbol{\alpha}'$.
It is the equivalent for WCNs to the conditional probability $P(k'|k)$
measuring the \textit{topological} correlations between nearest neighbors
\cite{Pastor-Satorras:2001}, but now with the extra information provided by
the dependence on strength and disparity. The conditional probability
$Q(\boldsymbol{\alpha}' | \boldsymbol{\alpha})$ measures the probability that,
when randomly choosing a vertex in the class $\boldsymbol{\alpha}$ and
following one of its edges with probability proportional to its weight, the
vertex at the other end belongs to the class $\boldsymbol{\alpha}'$. It is a
pure measure for WCNs, relating the effect of the weights to the strength of
the correlations. In a similar fashion as it is done in the case of unweighted
networks, we can define as a more practical correlation function, the average
degree of the neighbors of vertices of degree $\boldsymbol{\alpha}$, but now
weighted by the conditional probability $Q(\boldsymbol{\alpha}' |
\boldsymbol{\alpha})$, that is, $ \bar{k}_{nn}^{w}(\boldsymbol{\alpha}) =
\sum_{\boldsymbol{\alpha}'} k' Q(\boldsymbol{\alpha}' | \boldsymbol{\alpha}).
$ This is still a three variables function which is difficult to analyze.
Therefore, we coarse grain the degrees of freedom corresponding to $s$ and $Y$
in the following way:
\begin{equation}
\hspace{-0.1cm}
\bar{k}_{nn}^{w}(k) = \sum_{s,Y}
\frac{P(\boldsymbol{\alpha})}{P(k)}\bar{k}_{nn}^{w}(\boldsymbol{\alpha}) =
\frac{1}{N_k} \sum_{i \in \mathcal{V}(k)}
\frac{1}{s_{i}}\sum_jw_{ij} k_j,
\end{equation}
where the last term defines the numerical implementation of this function. The
summation over $i$ involves all vertices with degree $k$, 
$\mathcal{V}(k)$, and $N_k$ is the number of vertices with that degree. We
note that this measure coincides with the one proposed in
Ref.~\cite{Barrat:2004b}.

Turning now to three-vertex correlations, they are fully characterized by the three
vertex conditional probability $Q(\boldsymbol{\alpha}', \boldsymbol{\alpha}'' |
\boldsymbol{\alpha})$, which measures the likelihood that a vertex
$\boldsymbol{\alpha}$ is simultaneously connected to vertices
$\boldsymbol{\alpha}'$ and $\boldsymbol{\alpha}''$ when the weights of both
connections are considered.  In unweighted networks, the information
about three-vertex correlations can be conveniently compacted in the
degree-dependent clustering coefficient $\bar{c}(k)$. Similarly, for
WCNs we can generalize a weighted clustering
coefficient as
$
  \bar{c}^{w}(\boldsymbol{\alpha})=\sum_{\boldsymbol{\alpha}', \boldsymbol{\alpha}''}
  Q(\boldsymbol{\alpha}', \boldsymbol{\alpha}'' |\boldsymbol{\alpha}) r_{\boldsymbol{\alpha}'  \boldsymbol{\alpha}''}^{\boldsymbol{\alpha}},
$
where $r_{\boldsymbol{\alpha}' \boldsymbol{\alpha}''}^{\boldsymbol{\alpha}}$ is the
probability that two vertices in the classes $\boldsymbol{\alpha}'$ and
$\boldsymbol{\alpha}''$ are joined, provided that they have a common
neighbor in the class $\boldsymbol{\alpha}$. Once again, we can integrate
out the strength and disparity to obtain
\begin{equation}
  \bar{c}^{w}(k) = \sum_{s, Y}  \frac{P(\boldsymbol{\alpha})}{P(k)}
  \bar{c}^{w}(\boldsymbol{\alpha}), \label{eq:9}
\end{equation}
which represents the natural generalization for WCNs of
the clustering coefficient $\bar{c}(k)$.
Numerically, this function is given by
\begin{equation}
\bar{c}^{w}(k)=\frac{1}{N_{k}} \sum_{i \in \mathcal{V}(k)}
\frac{1}{s_{i}^2(1-Y_{i})}\sum_{jl} w_{ij}w_{il}a_{jl}.
\label{eq:10}
\end{equation}
Notice that this is different from the definition given in Ref.~\cite{Barrat:2004b}.

The zero measure of correlations is given by the so-called uncorrelated network ensemble, defined as the ensemble for which the joint distributions Eqs.~(\ref{eq:1}) factorize as $P(\boldsymbol{\alpha},
\boldsymbol{\alpha}')=kk'P(\boldsymbol{\alpha})P(\boldsymbol{\alpha}')/\langle k
\rangle^2$ and $Q(\boldsymbol{\alpha},
\boldsymbol{\alpha}')=ss'P(\boldsymbol{\alpha})P(\boldsymbol{\alpha}')/\langle s
\rangle^2$. In this case, one can easily prove that the measures defined above become
\begin{equation}
\bar{k}^{w}_{nn}(k) = \frac{ \langle k \bar{s}(k)
    \rangle}{\langle s \rangle} \quad \mathrm{and} \quad \bar{c}^{w}(k) = \frac{ \langle (k-1) \bar{s}(k)
    \rangle^2}{\langle k \rangle^2 \langle s \rangle N} ,
\end{equation}
where $\bar{s}(k) = \sum_{s, Y} s P(\boldsymbol{\alpha})/P(k)$. We have also assumed
that, for randomly assembled networks, $Q(\boldsymbol{\alpha}',
\boldsymbol{\alpha}'' |\boldsymbol{\alpha}) = Q(\boldsymbol{\alpha}'|\boldsymbol{\alpha})
Q(\boldsymbol{\alpha}'' |\boldsymbol{\alpha})$. As one can see, all these
functions become independent of the degree, so that any non-trivial
dependence on $k$ will signal the presence of two- and three-vertex
correlations, respectively.

In fact, one can realize
that, for any WCN, the joint distributions
$P(\boldsymbol{\alpha}, \boldsymbol{\alpha}')$ and $Q(\boldsymbol{\alpha},\boldsymbol{\alpha}')$
cannot factorize except for large degrees. Consider, for
instance, vertices of degree $k=1$ and strength $s$. The neighbors
of such vertices must have a strength that is, at least, $s$,
meaning that the properties of the neighbor depends on the
properties of the first vertex. Vertices of degree $k=2$ and
strength $s$, have weights in their connections that are a fraction
of $s$ and, then, the strength of their neighbors should be, at
least, the same fraction of $s$. The same effect is present, although in a
weaker form, for vertices of higher degrees. Therefore, {\it purely
uncorrelated WCNs cannot exist}. Just in the case of
large degrees, this structural correlations become very weak.

The highest level of randomness attainable in WCN
does not correspond  to the factorization of Eqs.~(\ref{eq:1})---which is
impossible---but of their marginal distributions, $P(k, k') =
\sum_{s, Y, s', Y'} P(\boldsymbol{\alpha}, \boldsymbol{\alpha}')$ and $Q(k, k') =
\sum_{s, Y, s', Y'} Q(\boldsymbol{\alpha},\boldsymbol{\alpha}')$. We can then define the corresponding conditional probability
$Q(k'|k)=\langle s \rangle Q(k,k')/\bar{s}(k)P(k)$ and the
two-vertex correlation function $\bar{k}_{nn}^{w, ns}(k) =
\sum_{k'} k'Q(k' | k)$, which filters out the structural correlations. It is numerically computed as
\begin{equation}
  \bar{k}^{w, ns}_{nn}(k)\equiv \frac{1}{N_k} \sum_{i \in \mathcal{V}(k)}
  \frac{1}{\bar{s}(k)}\sum_jw_{ij} k_j . \label{eq:12}
\end{equation}
In this function, the
contribution of every vertex $i$ depends on the average strength of
all the vertices with the same degree $k$. This implies an averaging that cancels out the effect of weight
induced correlations and yields a constant behavior when the marginal distributions factorize. The same line of reasoning also applies to clustering. The non-structural weighted clustering coefficient reads
\begin{equation}
\bar{c}^{w,ns}(k)=\frac{1}{N_{k}} \sum_{i \in k}
\frac{1}{\overline{s^2(1-Y)}(k)}\sum_{jl} w_{ij}w_{il}a_{jl},
\end{equation}
$\overline{s^2(1-Y)}(k)$ being an average over vertices of degree
$k$.

To check the accuracy of this approach, we need a null model as a
gauge for the presence or absence of non-structural correlations. This will imply
the construction of maximally random WCNs, which can be
easily inferred from the proposed formalism. The strategy consists in defining an ensemble at the hidden level where the local properties are fixed and where we can assume that the fundamental functions factorize~\cite{Boguna:2003b}. Instead of working with
the joint distributions, it is more convenient to define the new
quantities $r_{\boldsymbol{\alpha}, \boldsymbol{\alpha}'}$ and
$\bar{w}_{\boldsymbol{\alpha}, \boldsymbol{\alpha}'}$ ,
\begin{equation}
  r_{\boldsymbol{\alpha}, \boldsymbol{\alpha}'}=
        \frac{\langle k \rangle P(\boldsymbol{\alpha}, \boldsymbol{\alpha}')}{N
      P(\boldsymbol{\alpha})P(\boldsymbol{\alpha}' )}
, \qquad
 \bar{w}_{\boldsymbol{\alpha}, \boldsymbol{\alpha}'}=\frac{\langle s
\rangle Q(\boldsymbol{\alpha}, \boldsymbol{\alpha}')}{\langle k \rangle
P(\boldsymbol{\alpha}, \boldsymbol{\alpha}')}.
\label{eq:4}
\end{equation}
The first specifies the ratio between the number of connections among two classes and its
maximum possible number. The second corresponds to the average weight of an edge connecting two equivalence classes. Now, assuming the factorization of the fundamental functions, $r_{\boldsymbol{\alpha}, \boldsymbol{\alpha}'}$ and $\bar{w}_{\boldsymbol{\alpha}, \boldsymbol{\alpha}'}$ take the simple forms
\begin{equation}
  r_{\boldsymbol{\alpha}, \boldsymbol{\alpha}'}=\frac{kk'}{\langle k \rangle N}, \qquad
  \bar{w}_{\boldsymbol{\alpha}, \boldsymbol{\alpha}'}=\frac{\langle k \rangle s
    s'}{\langle s \rangle k k'},
\label{WeightsUncorr}
\end{equation}
a result implying that the topology of the network at the hidden level is decoupled from
the weights and, more importantly, independent of the disparity.
Using this result, we can generate a WCN without
two-point correlations (other than the structural ones) in the following way: we first construct an uncorrelated network with a given degree distribution $P(k)$ using
any of the algorithms available in the
literature~\cite{Boguna:2003b,Catanzaro:2005}. After the network has
been assembled, we assign an expected strength to each vertex
according to the distribution $g(s|k)$, under the constraint that
$P(k,s)=P(k) g(s|k)$. Finally, each edge is assigned a weight
according to Eq.~(\ref{WeightsUncorr}). In this way, we can generate
WCNs with any non-trivial correlation between strength
and degree and any form of the degree distribution. It is important
to notice that, in principle, the expected and final strength of a
vertex are not equal. However, one can prove that both quantities
converge on average.
\begin{figure}[t]
  \epsfig{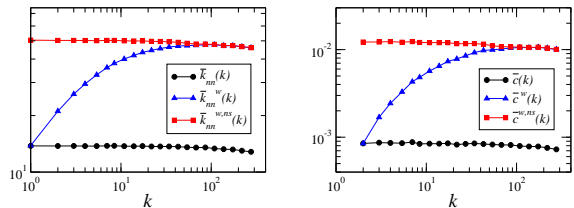} \caption{Correlation measures for a
  random WCN generated with the algorithm defined in the text with $P(k)
  \sim k^{-2.5}$ and $\bar{s}(k)\propto k^{1.5}$.} \label{fig:random}
\end{figure}

In Fig.\ref{fig:random}, we compare the
weighted correlation functions with their unweighted counterparts
for a WCN constructed with the algorithm explained above. We observe that the
weighted correlation functions are not flat, as they should be for
an uncorrelated network, but show a degree dependence for small $k$,
saturating to a constant plateau for large $k$. In contrast, the
non-structural functions recover the expected uncorrelated behavior
independent of $k$.

\begin{figure}[t]
  \epsfig{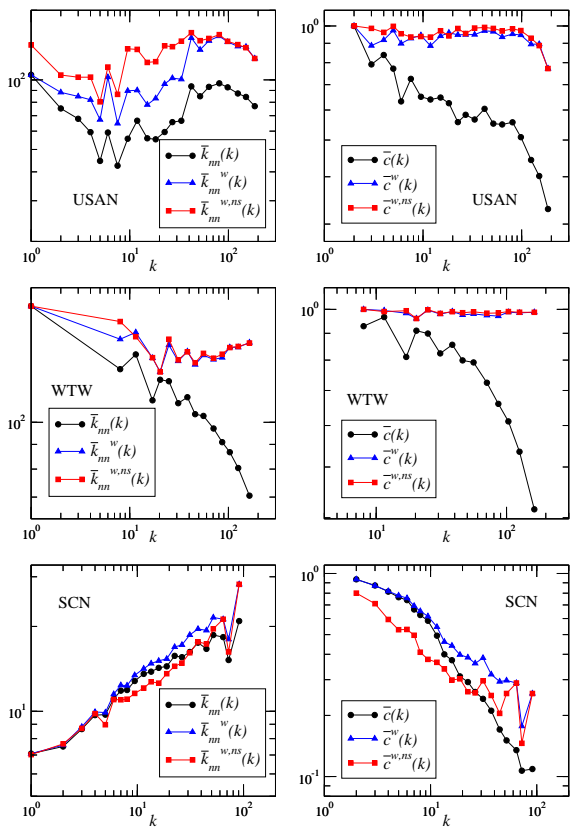} \caption{Correlation measures for
  real networks. From top to bottom, the US airport network (USAN) for the
  year 2005, the world trade web (WTW) for the year 2000, and the scientific
  collaboration network (SCN).} \label{fig:tricorr}
\end{figure}
Correlation measures for three different real networks are shown in
Fig.\ref{fig:tricorr}. The first observation is that, in general, weighted
measures greatly disagree with the unweighted ones, offering a completely
different picture with respect the the bare topology. For the USAN and the
WTW, the almost flat behavior proves that weighted two- and three-point
correlations are extremely weak, in contrast to the unweighted measures which
show important dependencies on $k$. This suggests that the understanding of
their formation processes or their modeling can be simplified by avoiding
correlations at the weighted level. Besides, the noticeable difference between
the weighted measures and its non-structural counterparts in the USAN graphs
manifests that structural correlations are more important for this network. On
the other hand, all measures follow a similar behavior in the SCN. However,
whereas the weighted two-point measure tells that the network is more
assortative than the unweighted estimation, the non-structural measure
indicates that this is due to an structural effect since, except for very high
degrees, $\bar{k}_{nn}^{w,ns}(k)<\bar{k}_{nn}(k)<\bar{k}_{nn}^{w}(k)$. This
effect is even more evident in the case of clustering. The weighted measure
proves that the tendency to form triangles is more important when weights are
considered. However, the non-structural measure is significantly smaller that
the unweighted one, which means that, when discounting structural effects, the
tendency to form triangles is in fact less pronounced.

Summarizing, we have shown that strict
uncorrelated WCNs at the local level do not exist due to the presence
of structural constraints. From a rigorous formal framework, we have defined
the appropriate weighted correlation measures that quantify the overall level
of correlations. We also propose complementary non-structural measures that
filter out the structural component and quantify the level of correlations in
the network as compared with the maximum randomness attainable. At this
respect, we have introduced an algorithm that generates maximally random WCNs
with an arbitrary $P(k,s)$ to be used as null models. We have applied our
formalism to analyze three different heterogeneous networks. The results make
evident the importance of taking into account weights to properly describe
this class of systems.

\begin{acknowledgments}
We thank M. Barth\'elemy for assistance with the USAN data and helpful
discussions, M.E.J. Newman for providing the SCN data, and A. Vespignani for
useful comments. M. B. acknowledges financial support from DGES, Grant No.
FIS2004-05923-CO2-02 and Generalitat de Catalunya Grant No. SGR00889. R.P.-S.
acknowledges financial support from the Spanish MEC (FEDER), under project No.
FIS2004-05923-C02-01 and additional support from the DURSI (Generalitat de
Catalunya, Spain).
  
\end{acknowledgments}


\end{document}